\documentclass[a4paper,fleqn,usenatbib]{mnras}

\usepackage{mathptmx}

\usepackage[T1]{fontenc}
\usepackage{ae,aecompl}

\usepackage{graphicx}	
\usepackage{amsmath}	
\usepackage{amssymb}	
\usepackage{epsfig}
\usepackage{pdflscape}

\def\msun{\hbox{M$_\odot$}}

\def\t4{\hbox{t$_{\rm 4}$}}

\def\cm3{\hbox{cm$^{-3}$}}

\input psfig.sty
\input epsf.sty

\title[MPs in Lindsay 1]{Evidence for multiple populations in the intermediate age cluster Lindsay 1 in the SMC\thanks{Based on observations made with ESO telescopes at the La Silla Paranal Observatory under Programme ID 096.B-0618(B).}}

\author[Hollyhead et al.]{
K. Hollyhead$^{1}$\thanks{E-mail: k.hollyhead@2013.ljmu.ac.uk)},
N. Kacharov$^{2}$,
C. Lardo$^{1}$,
N. Bastian$^{1}$,
M. Hilker$^{3}$,
M. Rejkuba$^{3,4}$,
\newauthor A. Koch$^{5}$,
E. K. Grebel$^{6}$ and
I. Georgiev$^{2}$
\\
$^{1}$Astrophysics Research Institute, Liverpool John Moores University, 146 Brownlow Hill, Liverpool L3 5RF, UK\\
$^{2}$Max-Planck-Institut f\"{u}r Astronomie, K\"{o}nigstuhl 17, D-69117 Heidelberg, Germany\\
$^{3}$European Southern Observatory, Karl-Schwarzschild-Stra\ss{}e 2, D-85748 Garching bei M\"{u}nchen, Germany\\
$^{4}$Excellence Cluster Universe, Boltzmannstr. 2, 85748, Garching, Germany\\
$^{5}$Department of Physics, Lancaster University, Lancaster LA1 4YB, UK\\
$^{6}$Astronomisches Rechen-Institut, Zentrum f\"{u}r Astronomie der Universit\"{a}t Heidelberg, M\"{o}nchhofstr. 12-14, 69120 Heidelberg, Germany\\
}

\date{Accepted XXX. Received YYY; in original form ZZZ}

\pubyear{2016}

\begin{document}
\label{firstpage}
\pagerange{\pageref{firstpage}--\pageref{lastpage}}
\maketitle

\begin{abstract}
Lindsay 1 is an intermediate age ($\approx 8$ Gyr) massive cluster in the Small Magellanic Cloud (SMC). Using VLT FORS2 spectra of 16 probable cluster members on the lower RGB of the cluster, we measure CN and CH band strengths (at $\simeq 3883$ and $4300 $ \AA~respectively), along with carbon and nitrogen abundances and find that a sub-population of stars has significant nitrogen enrichment. A lack of spread in carbon abundances excludes evolutionary mixing as the source of this enrichment, so we conclude that this is evidence of multiple populations. Therefore, L1 is the youngest cluster to show such variations, implying that the process triggering the onset of multiple populations must operate until at least redshift $\sim 1$.  
\end{abstract}




\section{Introduction}
\label{sec:intro}

Globular clusters (GCs) have been found to host multiple populations (MPs) of stars, an indication that they are not the simple stellar systems once thought to be. These MPs are characterised by abundance variations between stars that can be seen in both photometry (from splits and spreads in the Main Sequence (MS) or Red Giant Branch (RGB) in appropriate filters, for example (e.g. \citet{piotto13})) and spectroscopy (from chemical abundance anti-correlations (e.g. \citet{gratton12b})). Abundance variations have been observed for light elements such as C, N, O, Na, Al and Mg, which are often paired in anti-correlations. Iron has been found to exhibit very little variation in these clusters, which would be expected if supernovae were not responsible for the enrichment of the second population of stars.

Old GCs (>10 Gyr) in both the Milky Way and the LMC have been found to host MPs \citep{mucciarelli09, mateluna12}, while similarly aged, less massive open clusters have not. Therefore, it was traditionally thought that the main cluster property contributing to the presence, or lack thereof, of MPs is the cluster's mass. However, this has recently been called into question as younger ($\sim 1.5$ Gyr) clusters of comparable mass to GCs do not appear to host them (e.g. NGC 1806; \citet{mucciarelli14}). This indicates that age could also be a controlling parameter, and is what we aim to test here. 


The differences between the populations in younger clusters and old GCs is crucial for GC formation scenarios. It is currently accepted that the enrichment process must be external to the stars, as main sequence stars are also found to exhibit abundance variations, excluding processes such as evolutionary mixing within stellar interiors \citep{harbeck03}. 

Many scenarios describing the origin of MPs invoke multiple generations of stars in order to explain said features in the Colour Magnitude Diagrams (CMDs) of clusters and chemical variations. The AGB \citep[e.g.][]{dercole08}, FRMS (Fast Rotating Massive Star; e.g. \citet{decressin07} and interacting massive binaries \citep[e.g.][]{demink09} scenarios use the ejecta of evolved stars to pollute a second generation of stars forming later than the first, with age spreads of up to $\sim 300$ Myr. The early disc accretion scenario alternatively uses the ejecta of stars of the same generation to pollute pre-MS stars \citep{bastian13}. However, these scenarios cannot reproduce all light element abundance variations \citep{bastian15b} without succumbing to significant issues, such as the mass budget problem \citep[e.g.][]{larsen12,bastianlardo15,kruijssen15}. Additionally, sufficient gas reservoirs have not been found in Young Massive Clusters (YMCs) at the ages required for formation of a second generation with the proposed age spreads \citep{ivan15, longmore15}. 

The gap between the single population 1-3 Gyr clusters and those >10 Gyr with MPs is therefore an important age range to study to provide insight into exactly when and how MPs originate. This is the aim of this study into Lindsay 1 (hereafter L1), an intermediate age cluster in the SMC. At $\approx 8$ Gyr \citep{mighell98, glatt08}, it fits well into the unexplored age region for GCs. Additionally, it has low metallicity ([Fe/H] $\approx -1.35$, \citet{mighell98}) and is massive \citep[$\approx 1.7 - 2.6\times 10^5$ \msun ;][]{glatt11}, so is comparable with old GCs. In this study we obtain CN and CH band strengths \citep{kayser08,pancino10, lardo12} and C and N abundances to look for MPs in L1. 

In \S~\ref{sec:data} we discuss our data and its reduction, \S~\ref{sec:index} our measurement of CN and CH band strengths and C and N abundances, \S~\ref{sec:refine} discusses our determination of cluster members and \S~\ref{sec:results} and \S~\ref{sec:discussion} our results and discussion respectively.                 

\vspace{-0.6cm}

\section{Observations and data reduction}
\label{sec:data}

Our spectroscopic data for L1 was obtained in one observing run on 06/10/15 at the ESO VLT telescope based in Paranal, Chile using the MXU mode on the FORS2 spectrograph. We took five science exposures along with bias frames, flat fields and an arc lamp spectrum for wavelength calibration. Our instrument configuration consisted of the GRIS 600B+22 grism with two 2kx4k E2V CCDs ($15~\mu m$ pixel size), mosaic-ed, which are sensitive in the blue range below 600 nm, a requirement to accurately measure CN and CH bands at $\simeq 3883$ and 4300 \AA~respectively. We centred the cluster on the master chip (hereafter chip 1), while the slave chip (chip 2) sampled outer regions of the cluster. The resolution of our spectra is $R = \lambda/\delta\lambda \simeq 800$ and the spectral range covered by most stars was $\sim 3300 - 6600$ \AA.

Archival pre-imaging in the V and I bands was already available for this cluster (ESO-programme: 082.B-0505(A) P.I. D. Geisler), which was used to select targets, with an aim to sampling primarily the lower RGB. 34 targets were chosen across chips 1 and 2, though the 1" slit width prevented us from selecting targets in the central regions of the cluster due to crowding. Where it was impossible to image a primary target due to slit positioning, a random star was chosen in its place. 

The spectra were reduced using tools available in the {\sc iraf} software package \citep{iraf}. We combined bias frames and subtracted this from all images followed by combining and normalising flat fields and applying this to the science and arc lamp frames. Cosmic rays were removed using the L.A.Cosmic {\sc iraf} routine \citep{lacos}. When we first traced the apertures across the images, we found very little curvature and so decided not to use a correction routine but allow the {\it apall} task to account for this when extracting the spectra. We opted to first combine the five science exposures and then extract the spectra for all targets using the {\it apall} task. The {\it identify} and {\it dispcor} tasks were then used to wavelength calibrate the spectra using the arc lamp exposure.

\begin{figure}
 \includegraphics[width=8cm]{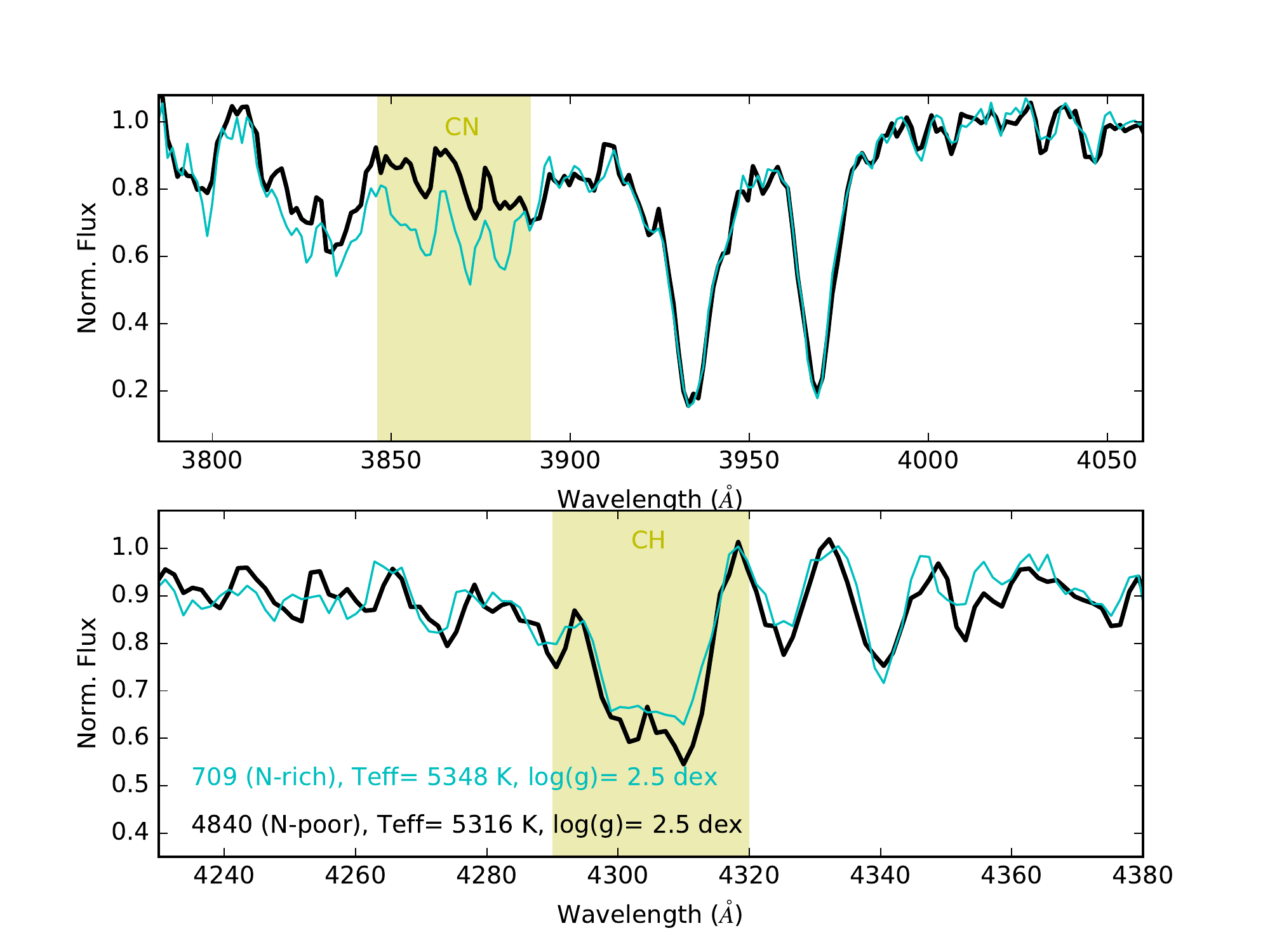}
 \caption{Samples of the final extracted normalised spectra showing the CN and CH bands of two stars in the top and bottom plots respectively.}
 \label{fig:spectra}
 \end{figure}

Fig.~\ref{fig:spectra} shows sections of the spectra of two stars with nearly identical $T_{eff}$ and log(g) around the CN and CH molecular absorption bands to illustrate the quality of our final spectra. Differences in the strength of the CN absorption can easily be seen, while the CH band is approximately the same. 

 \vspace{-0.6cm}

\section{Index measurement and abundance analysis}
\label{sec:index} 

\subsection{CN and CH band strengths}
\label{sec:cnch}

We measured a set of indices quantifying the strength of the $UV$ CN band (S$\lambda3883$) and the G band of CH (CH$\lambda4300$), the CaII (H+K) and Fe5270 lines. We adopted the same indices as defined in \citet{Norris81,Worthey94}, and \citet{Lardo13}. The uncertainties related to the index measurements are obtained with the expression derived by \citet{Vollmann06}. The indices, together with other useful information on the target stars, are listed in Table~\ref{tab:sample}. The measurement of the CaII (H+K) and Fe5270 lines were used to help determine cluster membership, as discussed in \S~\ref{sec:refine}. The typical S/N for the CN and CH bands used for calculations is $\sim25$ and $\sim40$ respectively.

\vspace{-0.5cm} 

\subsection{C and N abundances}
\label{sec:cn}

To obtain a quantitative estimate of the detected enhancements in nitrogen of L1 stars, we derived  [C/Fe] and [N/Fe] abundance ratios via spectral synthesis of the CH band at $\simeq$4300~\AA~and the CN bands at $\simeq$3883~\AA, respectively.

The atomic and molecular line lists were taken from the latest Kurucz compilation downloaded from F. Castelli's website\footnote{\url{http://wwwuser.oats.inaf.it/castelli/linelists.html}} both for atomic and molecular transitions. Model atmospheres were calculated with the ATLAS9 code starting from the grid of models available in F. Castelli's website using the appropriate values of [Fe/H], $T_{\rm{eff}}$, $\log$(g), and v$_{\rm mic}$. 

The effective temperature, $T_{\rm{eff}}$, was calculated using the \citet{Alonso99} $T_{\rm{eff}}$-colour calibration. We used the $(V-I)$ colour adopting $E(B-V) =0.06$ and [Fe/H] =-1.35 from \citet{mighell98}. The surface gravity was determined using $T_{\rm{eff}}$, a distance modulus of $(m-M)_{V}$=18.88 \citep{sersic62}, bolometric corrections BC(V) from \citet{Alonso99} and assuming a mass of 0.95 $\msun$. We assigned microturbulent velocity $v_{t}=2.0 \, kms^{-1}$ to all our stars. Table~\ref{tab:sample} lists the adopted atmospheric parameters and their associated uncertainties.

Model spectra have been computed by means of SYNTHE code developed by Kurucz. Abundances have been derived through a $\chi^{2}$ minimisation between the observed spectrum and a grid of synthetic spectra calculated at different abundances. Abundances for C and N were determined together in an iterative way, assuming the measured C abundance to derive N from the synthesis of the molecular CN band. The adopted solar abundances are from \citet{Asplund09}.

The sensitivity of the derived [C/Fe] and [N/Fe] abundances to the adopted atmospheric parameters where determined in the same fashion as done in \citet{Lardo13}. Briefly, we repeated our abundance analysis by changing only one parameter at each iteration for three stars that are representative of the temperature and gravity range explored. Typically for the temperature, we found $\delta$[C/Fe] / $\delta T_{eff} \simeq$ 0.12 dex and $\delta$[N/Fe] / $\delta T_{eff} \simeq$ 0.15 dex. The errors in abundances are mostly due to uncertainties in gravity and those due to microturbulent velocity are negligible (on the order of 0.05 dex or less). In addition to the stellar parameters and oxygen abundance errors, an uncertainty exists in the measurement of the individual abundances. This intrinsic error was estimated by means of Monte Carlo simulations by repeating the fitting procedure using a sample of 500 synthetic spectra where Poissonian noise has been injected in order to reproduce the noise conditions observed around the analysed bands. These uncertainties are of the order of $\simeq$0.15 and $\simeq$ 0.25 dex, respectively for  C and N. All these sources of error were combined to give the final errors on [C/Fe] and [N/Fe] abundances ($\sim$0.2 and 0.3 dex, respectively).


\begin{table*}
 \centering
 Stellar properties
 \begin{tabular}{c c c c c c c c c c c}
 StarID &  & S$\lambda$3883 & CH$\lambda$4300 & [C/Fe] & [N/Fe] & Fe5270 & CII (H+K) & Teff & log(g) & RV \\
 \hline
0098  &   ... & -0.18$\pm$     0.05 &     -0.44$\pm$     0.05 &      0.15$\pm$     0.20 &      0.13$\pm$     0.30 &     -0.18$\pm$     0.02 &     25.51$\pm$     7.45 & 5448$\pm$97 &      2.80$\pm$     0.04 &     146.3\\
0229 &  ... &  -0.17$\pm$     0.05 &     -0.41$\pm$     0.05 &      0.16$\pm$     0.20 &      0.06$\pm$     0.30 &     -0.17$\pm$     0.02 &     27.11$\pm$     5.02 & 5327$\pm$92 &      2.50$\pm$     0.04 &     140.0\\
0384 & ...  &     -0.14$\pm$     0.07 &     -0.45$\pm$     0.08 &      0.08$\pm$     0.20 &      0.58$\pm$     0.30 &     -0.18$\pm$     0.02 &     22.60$\pm$     8.37 & 5544$\pm$101 &      3.00$\pm$     0.04 &     158.1\\
0511 & ... &     -0.05$\pm$     0.04 &     -0.43$\pm$     0.06 &      0.18$\pm$     0.20 &      1.02$\pm$     0.30 &     -0.17$\pm$     0.02 &     26.23$\pm$     8.38 & 5495$\pm$99 &      2.80$\pm$     0.04 &     144.5\\
0709 & ... &      0.11$\pm$     0.05 &     -0.44$\pm$     0.05 &      0.01$\pm$     0.20 &      1.28$\pm$     0.30 &     -0.17$\pm$     0.02 &     27.16$\pm$     5.52 & 5348$\pm$93 &      2.50$\pm$     0.04 &     149.8\\

\end{tabular}
\caption{Sample of derived properties for some targets. The full table for all 34 sources is available in the supplementary material online.}
\label{tab:sample}
\end{table*}

\vspace{-0.6cm}

 \section{Membership determination}
 \label{sec:refine}
 
  \begin{figure}
  \includegraphics[width=7.5cm]{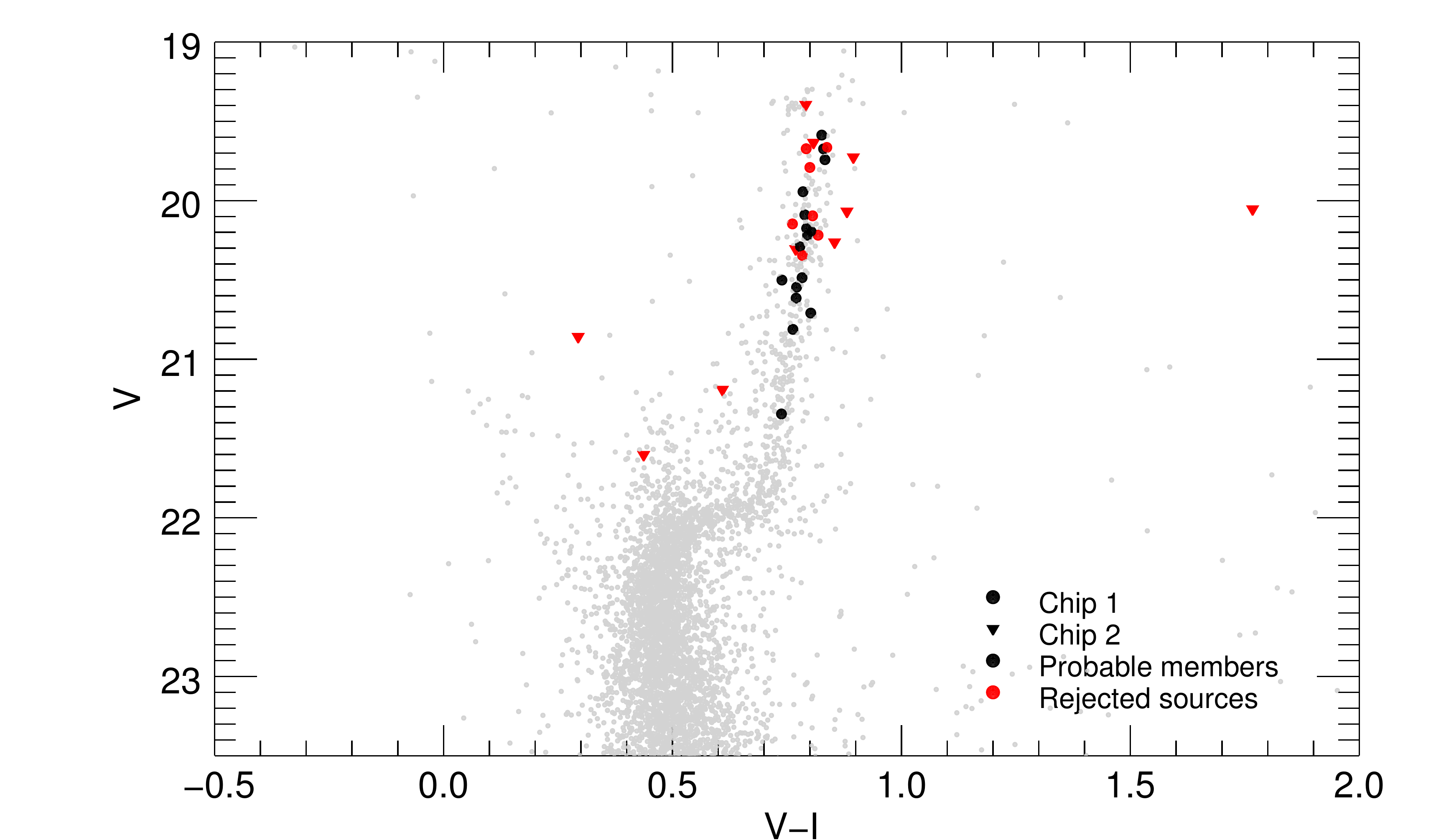}
  \includegraphics[width=7.5cm]{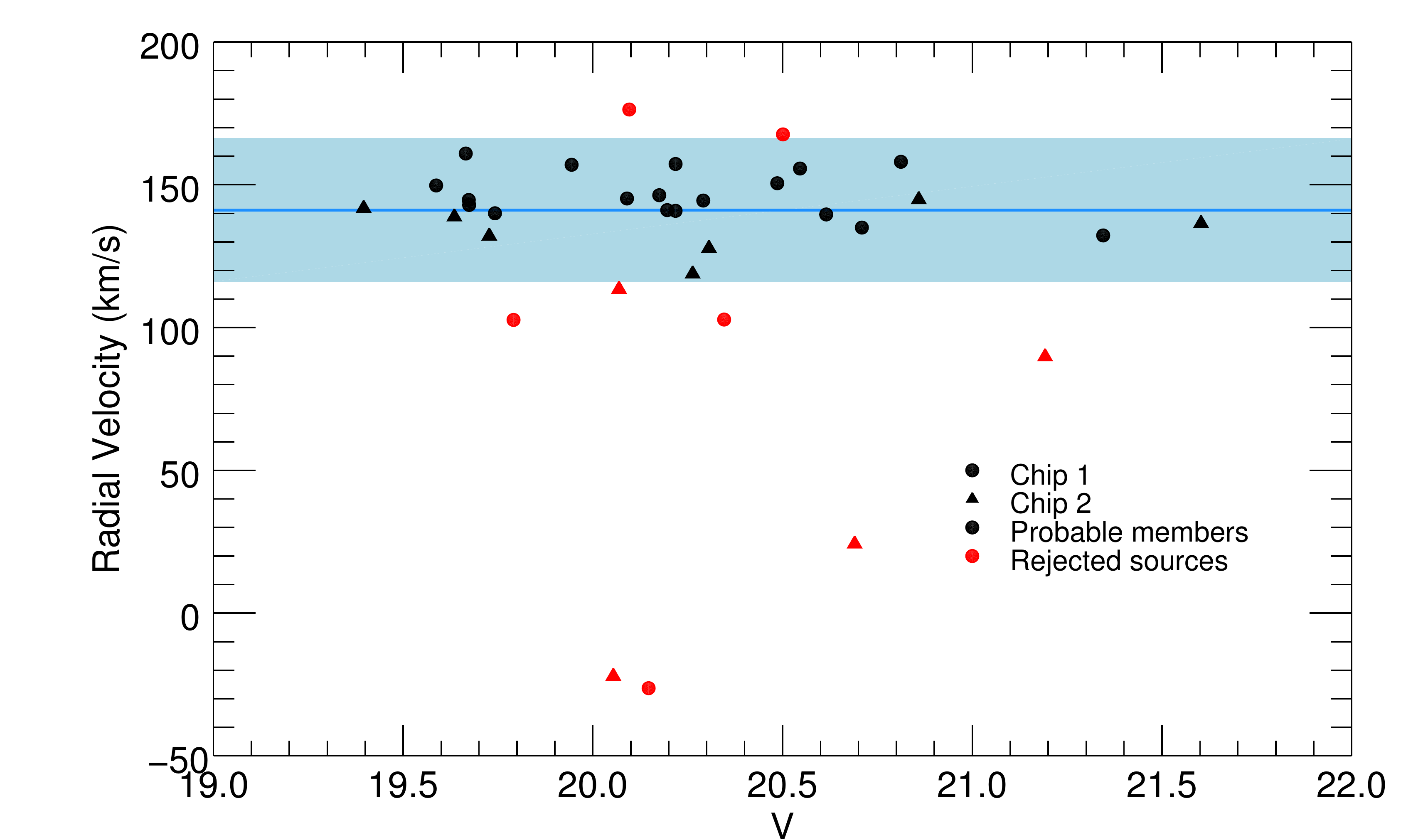}
  \includegraphics[width=7.5cm]{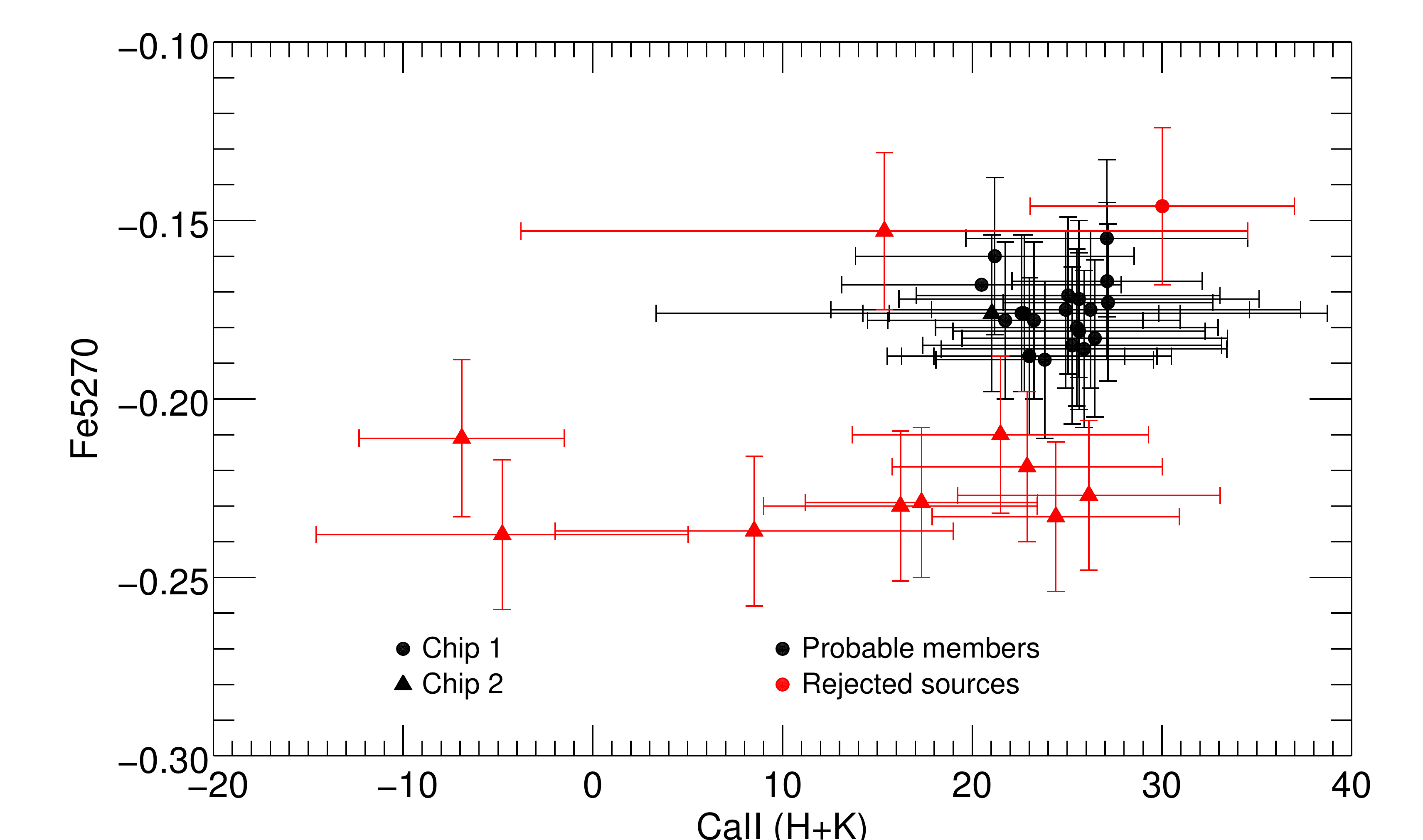}
  \caption{Criteria used to determine true cluster members: the colour magnitude diagram for all stars in the pre-imaging catalogue in grey with the 34 target stars overplotted, radial velocities for the targets in our sample and CaII (H+K) and Fe5270 abundances plotted against each other for all 34 sources. Chip 1 stars are circles, while chip 2 are triangles and membership is demonstrated by colour, with black representing members and red non-members. }
  \label{fig:members}
  \end{figure}
 
 An important step in ensuring our C/N and CH/CN distributions and any implications are accurate is discerning the true cluster members from contaminants. We utilised several different methods besides the traditional radial velocity (RV) measurements as we have fairly large errors in RVs for low resolution spectra. There is still a possibility of some field interlopers due to the potentially compatible field star RVs in the SMC, \citep{carrera08} however, compliance with all four membership criteria makes this very unlikely. Fig.~\ref{fig:members} shows the plots for all of our membership tests. Red points indicate those targets identified as non-cluster members.   
 
 Firstly, we examined the CMD shown in Fig.~\ref{fig:members} for outliers. The points displayed in red are all those that were rejected, not just those identified from the CMD. Removed stars are those that lie far away from the RGB, which were likely random stars chosen to replace priority targets that could not be used due to slit positioning. Secondly, RVs were found using the {\it fxcor} utility in {\sc iraf} with a template spectrum of a likely member star, with a heliocentric-corrected velocity found independently from {\it rvidlines}. The RVs are listed in Table~\ref{tab:sample} and displayed in Fig.~\ref{fig:members} against the apparent V band magnitude. We measured the median velocity of all stars as 141.21 km/s, in agreement with the expected velocity for L1 stars of 145.3 km/s from FORS2 spectra by \citet{parisi15}. There may also be a small offset due to slits being placed slightly off-centre from the star. A reasonable error on RV measurements is $\sim 25$ km/s, which we estimated from running the {\it rvidlines} {\sc iraf} package across different sections of several spectra to gauge the difference between the bluer and redder regime and any sources that were more than 25 km/s away from our median were unlikely to be members. We then measured the CaII (H+K) and Fe5270 lines and any sources more than 1 $\sigma$ away from the median value of each were considered non-members.
 

  
 Finally, we visually inspected the spectra of the targets to ensure that they were all reliably RGB stars. We excluded stars 0008, 0101, 0140 and 0203 on this basis, as two had strong absorption features in the red part of the spectrum, suggesting the stars were too cool and the other two had strong hydrogen absorption lines indicating that they are probably hotter, unevolved MS stars.   

 Also we note that star 5059 displays a prominent Mg triplet (5177\AA) feature in its spectra, indicating that this star is possibly more metal-rich than the rest of the cluster population. Also, spectral synthesis  reveals that it has [C/Fe] abundance significantly higher ([C/Fe]=0.63$\pm$0.2 dex) than the bulk of L1 stars, for which we found [C/Fe]$\simeq$0.1 dex. We also exclude this star from the following analysis.
 
 These criteria eliminated 5 stars from the CMD, 9 stars with outlying RVs, 5 using CaII (H+K) and 12 using Fe5270 (though several of these overlap), which left us with 15 cluster members out of the 34 original targets. Non-members are indicated in Table~\ref{tab:sample} with their reason for exclusion.  

\vspace{-0.75cm}
 
 \section{Results}
 \label{sec:results}
 
  \begin{figure*}
  \includegraphics[width=7cm]{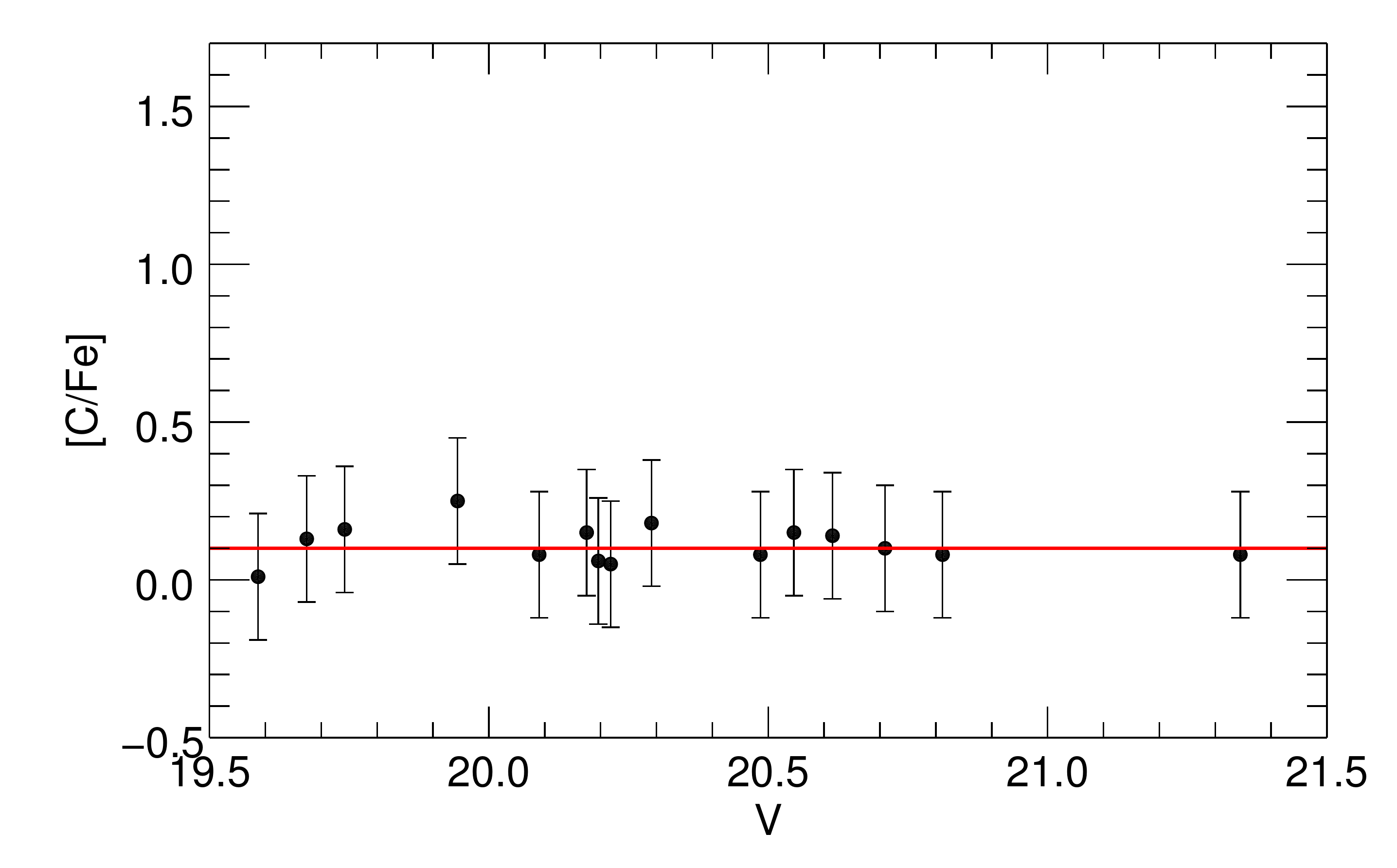}
  \includegraphics[width=7cm]{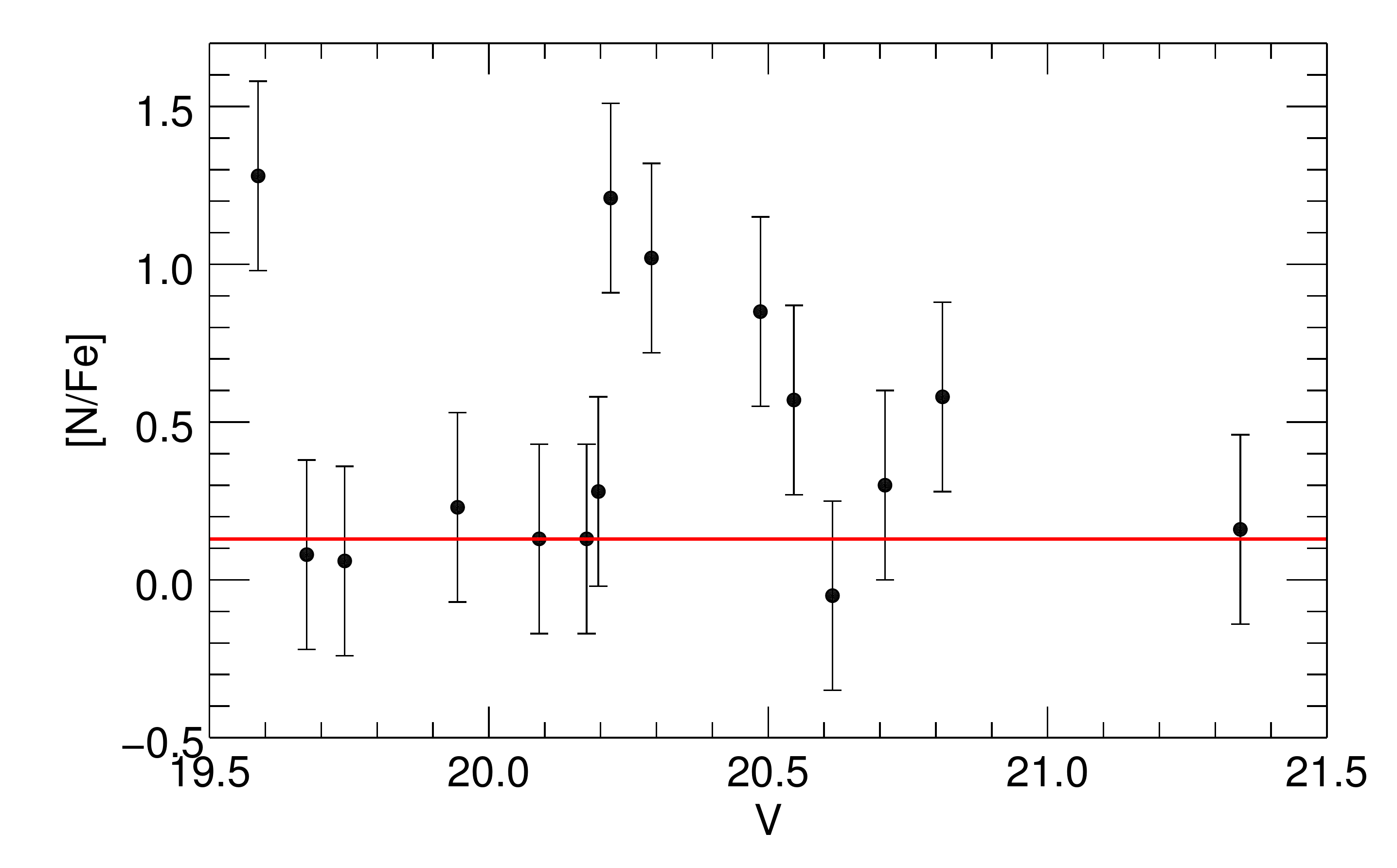}
  \caption{The left and right plots show carbon and nitrogen abundances ([C/Fe] and [N/Fe] respectively) for all true cluster members plotted against their apparent V band magnitude, with the median value of the main population shown as a red line.}
  \label{fig:cvnv}
  \end{figure*}
  
 In Fig.~\ref{fig:cvnv} we plot the C and N abundances against V band magnitude to check for effects of evolutionary mixing. The left plot shows negligible spread in [C/Fe], while the right plot shows a spread of $\sim 1$ dex in [N/Fe]. The y axes are of equal width to demonstrate the effect of the spread. The red lines are the median abundances for each calculated with all sources for C and the lower group of stars for N abundances. 
 
 \begin{figure}
 \includegraphics[width=7.5cm]{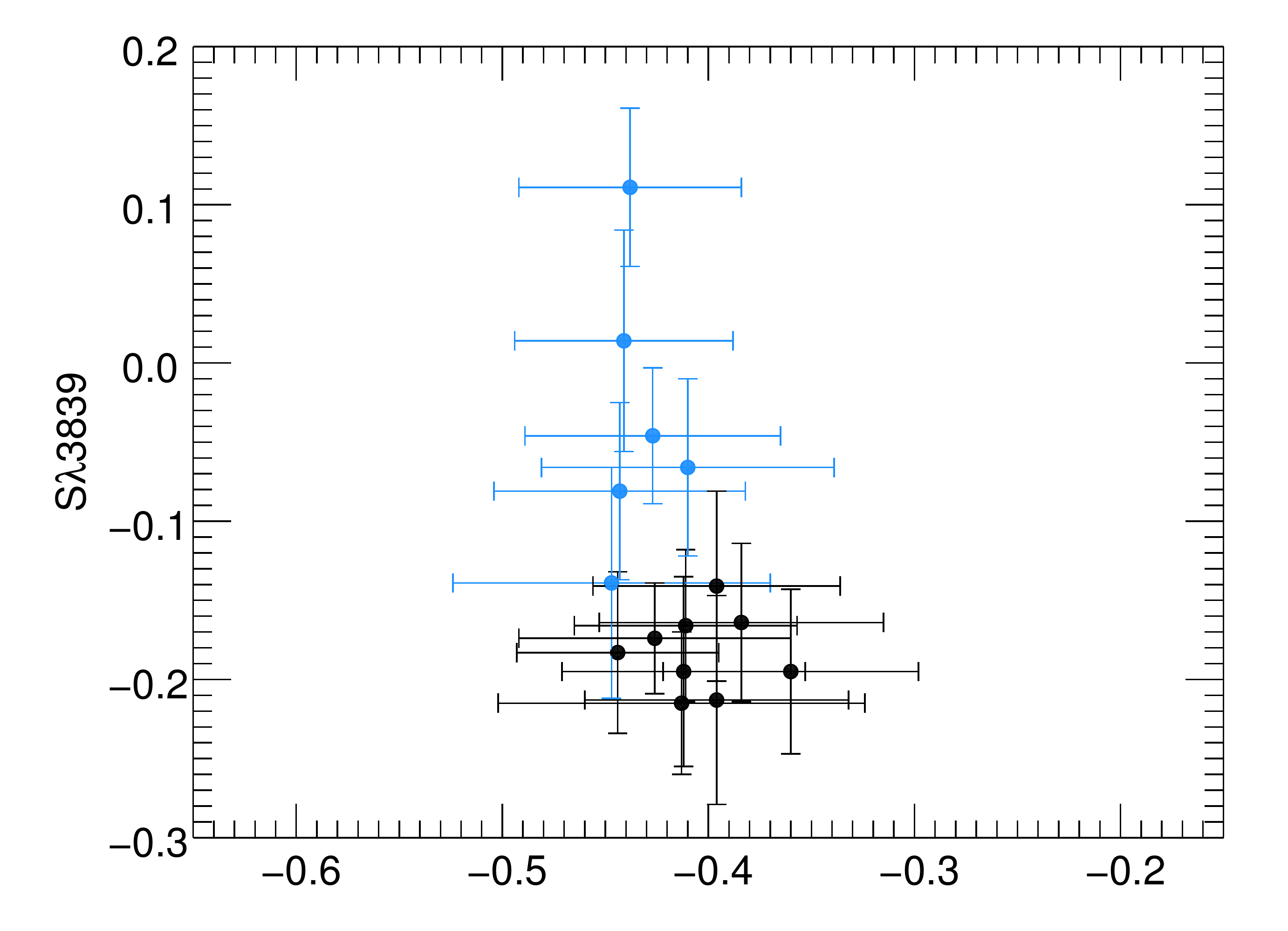}
 \includegraphics[width=7.5cm]{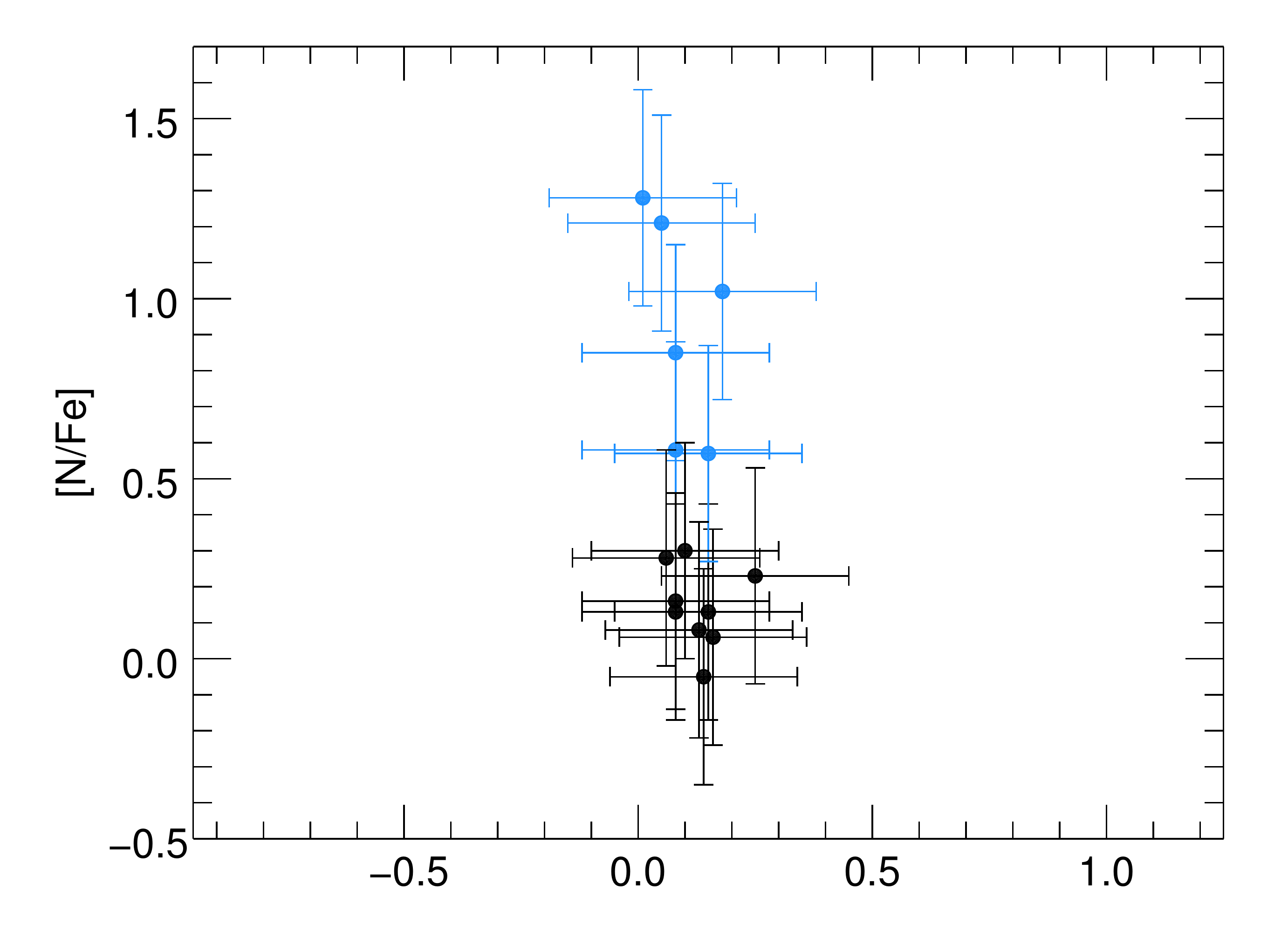}
 \caption{The top plot shows CN S$\lambda$3839 and CH$\lambda$4300 band strengths for all sources considered cluster members and the lower plot the analagous [N/Fe] against [C/Fe]. The blue points are the sources that could be considered nitrogen enriched and members of a second population in L1.}
 \label{fig:cnch}
 \end{figure}
 
 Fig.~\ref{fig:cnch} shows the band strengths, S$\lambda$3883 and CH$\lambda$4300 plotted for sources that were determined to be members of the cluster, with equally sized axes to emphasise the difference in spread. The plot shows negligible spread in the CH index (tracing carbon) within the error bars and therefore suggests that the sources all have the same amount of carbon. Conversely, the CN index (tracing nitrogen) shows a much larger spread that is significant compared to errors. This indicates that we have nitrogen rich stars present in the sample, and is strong evidence for the presence of MPs as seen in Milky Way, LMC and Fornax GCs (e.g. \citet{larsen14}). The sources shown in blue can be considered N-enhanced and were selected based on Fig.~\ref{fig:cvnv}. 
 
 The lower plot in Fig.~\ref{fig:cnch} shows C and N abundances. Again, no significant spread is seen in [C/Fe], while 6 of the 16 stars determined as final cluster members could be considered to have enriched levels of N. This spread in the N abundance is larger than errors, up to $\sim1$ dex.
 
\vspace{-0.6cm}
 
\section{Discussion and conclusions}
\label{sec:discussion}

We obtained spectroscopy for 34 targets towards L1, and have determined through various methods that 15 of these sources are true cluster members and reliably lower RGB stars. Out of these 15 stars, 6 show enriched [N/Fe] compared to a fairly constant [C/Fe] throughout all sources, which is evident in Fig.~\ref{fig:cnch}. This is strongly indicative of the presence of MPs in L1, the youngest cluster to show abundance variations, though not the least massive \citep[NGC 6362 at $5 \times 10^4$ \msun,][]{mucciarelli16}. This indicates that the unknown mechanism responsible for MPs operated more recently than previously thought, though sample size is currently very small and more intermediate age GC targets can be studied in the SMC. Niederhofer et al, in prep, have also found evidence in support of MPs in L1 using HST photometry. In certain filter combinations the RGB splits into two sequences, in support of our findings using FORS2 spectroscopy. Stars in our study that overlap with their catalogue also lie on the correct RGB branch depending on whether they are N-enriched or not.


Evolutionary mixing can be disregarded as an explanation for elevated N abundances, as the C abundances show no variation. This indicates that our abundance estimates for the main N-poor population should be similar to those of the stars' original gas cloud (apart from some small evolutionary effects due to internal stellar mixing). Additionally, as our sources are lower RGB stars, they are fainter than the bump in the LF \citep[$V_{BUMP} = 19.30 \pm 0.05$;][]{alcaino03} meaning that any evolutionary mixing should have had minimal impact on the C and N abundances \citep[e.g.][]{gratton00}. 

We have also visually examined the N-enhanced stars using HST ACS F555W band images (proposal ID 10396, P.I Gallagher) taken from the Hubble Legacy Archive (HLA). Though the ACS image was smaller than our coverage, 3 of the 6 enriched stars were within ACS field of view, including 0709, the most enriched star. All of these targets appeared to be reliable sources, isolated single stars without contamination from nearby objects. Based on evidence for a lack of evolutionary mixing, the quality of the sources and our stringent membership tests, we believe the spread in [N/Fe] to be real.

It is important to note, however that we are only sampling the outer regions of L1, as the centre of the cluster is too crowded to obtain spectra of single stars with the slits on FORS2. Therefore, the observed ratios (N-enriched/N-normal) cannot be used to derive $F_{enriched}$. This may have affected previous studies of young and intermediate age LMC clusters (e.g \citet{mucciarelli08,mucciarelli14}), hence, HST imaging of these clusters in filters sensitive to MPs should also be undertaken. In order to investigate other light elements, time-consuming, high resolution spectroscopy is necessary, however, our method can be regarded as a promising way to identify and study MPs on shorter timescales. 
 
\vspace{-0.5cm}

 




\bibliographystyle{mnras}
\bibliography{biblio} 





\bsp	
\label{lastpage}
\end{document}